# WST – Wide-field Spectroscopic Telescope: The Next Leap in Wide-field Spectroscopy.

Roland Bacon[1], Olga Bellido[2], Philippe Dierickx[1], Paolo Franzetti[6], Roelof S. de Jong[2], David Lee[7], Vincenzo Mainieri[3], Sofia Randich[11], Richard I. Anderson[15], Francesco Belfiore[3], Henri Boffin[3], Julia Bryant[4], Kjetil Dohlen[5], Richard Ellis[9], Gaston Gausachs[8], Alexandre Jeanneau[1], Laura Magrini[11], Simon O'Toole[23], Michele Moresco[22], Arlette Pécontal[1], Anna Puglisi[11,18], Paula Sanchez Saez[3], Rodolfo Smiljanic[14], Stefano Zibetti[11], Bodo Ziegler[10], Matthew Colless[8], Matt Lehnert[1], Pietro Schipani[13], Laurence Tresse[5], Joe Barrow[23], Nicolas Bouché[1], Dimitri Buffat[5], Vincenzo De Caprio[13], Letizia P. Cassarà[6], Younes Chahid[7], Corentin Cudennec[1], Bianca De Caro[6], Simone D'Auria[11], Laurane Fréour[10], Marco Fumana[6], Maria Cristina Fortuna[11], Adriana Gargiulo[6], Pascale Hibon[3], Alessandro Meoli[3], Matteo Munari[12], Aaron Omadutt[2], Dinko Milaković[21], Peder Norberg[19,20], Marco Palla[11], Sébastien Pernecker[24], Marco Scodeggio[6], Maxime Rombach[24], Will Saunders[23], Elmo Tempel[16], and Nicholas A. Walton[17].

[1] Centre de Recherche Astrophysique de Lyon, Lyon, France

[2] Leibniz Institute for Astrophysics Potsdam (AIP), Potsdam, Germany

[3] European Southern Observatory, Garching, Germany

[4] Astralis, Sydney Institute for Astronomy, School of Physics, University of Sydney, Sydney, Australia

[5] Aix Marseille Univ., CNRS, CNES, Laboratoire d'Astrophysique de Marseille, Marseille, France

[6] INAF–IASF Milano, Milan, Italy

[7] UK Astronomy Technology Centre, Edinburgh, United Kingdom

[8] Research School of Astronomy and Astrophysics, Australian National University, Canberra, Australia

[9] University College London, London, United Kingdom

[10] Department of Astrophysics, University of Vienna, Vienna, Austria

[11] INAF–Osservatorio Astrofisico di Arcetri, Florence, Italy

[12] INAF–Osservatorio Astrofisico di Catania, Catania, Italy

[13] INAF–Osservatorio Astronomico di Capodimonte, Naples, Italy

[14] Nicolaus Copernicus Astronomical Center, Polish Academy of Sciences, Warsaw, Poland

[15] Institut für Astrophysik und Geophysik, Georg-August-Universität Göttingen, Göttingen, Germany

[16] Tartu Observatory, University of Tartu, Tõravere, Estonia

[17] Institute of Astronomy, University of Cambridge, Cambridge, United Kingdom

[18] University of Southampton, Southampton, United Kingdom

[19] Institute for Computational Cosmology, Durham University, Durham, United Kingdom

[20] Centre for Extragalactic Astronomy, Durham University, Durham, United Kingdom

[21] INAF - Osservatorio Astronomico di Trieste, Trieste, Italy
[22] University of Bologna, Bologna, Italy
[23] Macquarie University, Sydney, Australia
[24] EPFL, Lausanne, Switzerland

**ABSTRACT**

The Wide-field Spectroscopic Telescope (WST) is a concept for a dedicated 12-m spectroscopic survey facility designed to address some of the most important questions in astrophysics in the 2040s. The WST will provide unprecedented spectroscopic survey capabilities by operating simultaneously over a 2-degree diameter field of view 54 low-resolution spectrographs fed by 30,000 fibres, 8–16 high-resolution spectrographs fed by 2000 fibres and a large panoramic low-resolution integral-field spectrograph.

Supported by Horizon Europe, the concept study has refined the science cases, facility architecture, operations model, sustainability strategy, and technology roadmap. The resulting reference design demonstrates that the WST is both scientifically transformative and technically feasible, while identifying the developments required to mitigate the remaining risks. The WST is designed as an ESO flagship facility for the post-ELT construction era and a key spectroscopic complement to the major imaging, time-domain, and multi-messenger facilities of the coming decades.

# 1. INTRODUCTION

The Wide-field Spectroscopic Telescope (WST) concept study aims to design a next-generation spectroscopic survey facility capable of obtaining accurate spectra for hundreds of millions of Galactic and extragalactic sources across large areas of the sky. By delivering unprecedented survey efficiency, multiplexing, and spectral capability, the WST will address key challenges in astrophysics, ranging from the formation of the Milky Way to the evolution of galaxies, and the nature of dark matter and dark energy.

The WST is being developed as a candidate for ESO's Expanding Horizons initiative [1] and will be a flagship facility for the post-ELT construction era. By delivering unprecedented spectroscopic survey capabilities, it will enable transformative science across a wide range of astrophysical disciplines and provide the essential spectroscopic complement to the major imaging and time-domain surveys of the coming decades.

The WST astrophysical research goals and the top-level facility capabilities derived from them [2], together with initial concepts for the telescope [3] and instruments [4], were presented at the 2024 SPIE conference in Yokohama. Since then, and with the support of Horizon Europe funding, the concept study has progressed significantly. Work has been carried out across all major aspects of the project, including the refinement of the science cases and top-level requirements, facility design and trade-off analyses, operations concepts, sustainability considerations, and cost and risk assessments.

This paper presents the current status of the WST concept study, with particular emphasis on the evolution of the facility design and the key architectural choices that have emerged from the ongoing trade-off analyses. More details of the telescope reference design and the instrumentation concepts can be found in [8] and [22], respectively.

# 2. EVOLUTION OF TOP LEVEL REQUIREMENTS

Table 1 presents the latest version of the top-level requirements. In comparison to the initial version presented in [2][1], several key changes have been implemented:

The MOS-LR multiplex has been expanded to 30,000 instead of 20,000, giving a multiplex nearly an order of magnitude higher than the best existing instruments. This substantial increase will enhance the scientific impact of the facility, enabling highly ambitious surveys to be conducted within a few years. Consequently, the WST will be a unique facility for large-scale optical surveys of stars and galaxies, as illustrated in Figure 1. This ambitious multiplex goal necessitates an increase in the complexity of the positioner, which now must position 50% more fibres within the same volume, requiring the development of new concepts, as discussed further in the system design section. Additionally, it requires an increase in the number of spectrographs, which has implications for the volume and cost.

The wavelength range of the MOS-LR and IFS systems has been restricted to 370-930 nm (previously 350-970 nm for the MOS-LR and 370-970 nm for the IFS). The blue wavelength cut of 370 nm for the MOS-LR is motivated by the critical loss of UV transmission of the fibre and the limited selection of transparent glasses below 370 nm, which contribute to the spectrograph costs. In the red region, between 930 and 970 nm, there are numerous sky absorption and emission lines, which, in conjunction with the anticipated lower detector quantum efficiency, restrict the performance of the system. The slightly restricted wavelength range allows a slight increase in spectral resolution, particularly in the red region.

Another significant enhancement is the incorporation of a ground-layer adaptive optics system (GLAO) within the IFS optical path. This system will exclusively utilize natural guide stars which will be extracted from a 6 arcminute aperture encompassing the 3x3 arcminute-squared science field (Figure 2). Due to the concurrent operation of the IFS and the MOS, the use of sodium laser guide stars was deemed unsuitable as they could potentially interfere with the MOS operation. Nevertheless, even in the absence of laser guide stars, performance simulations indicate that 80% sky coverage is achieved at the galactic pole. Furthermore, considering the Paranal turbulence statistics, GLAO will yield 0.5 arcsecond image quality half the time, compared to only 20% without GLAO correction. This enhancement will positively impact IFS performance on resolved sources and significantly increase the survey speed for point sources.

*Table 1: Top Level Requirements*

| | |
|---|---|
| Telescope Aperture | 12 m (100 $m^2$) |
| Telescope FoV | 2 deg diam. (3.1 $deg^2$) |
| Tel. Spec. Range | 370-1600 nm |
| MOS LR Multiplex | 30,000 |
| MOS LR Resolution | 3,800 @ 420 nm 4,900 @ 680 nm |
| MOS LR Spec. Range | 370-930 nm (simultaneous) |
| MOS HR Multiplex | 2,000 |
| MOS HR Resolution | 40,000 |
| MOS HR Spec. Range | 401-443, 452-500, 542-600, 608-672 nm |
| IFS FoV & sampling | 3x3 $arcmin^2$ 0.25x0.25 $arcsec^2$ spaxel |
| IFS Resolution | 4,800 @ 480,750 nm |
| IFS Spec. Range | 370-930 nm (simultaneous) |
| IFS Patrol Field | 13 arcmin diameter |
| MOS & IFS parallel operation | |
| ToO implemented at telescope & fibre level | |

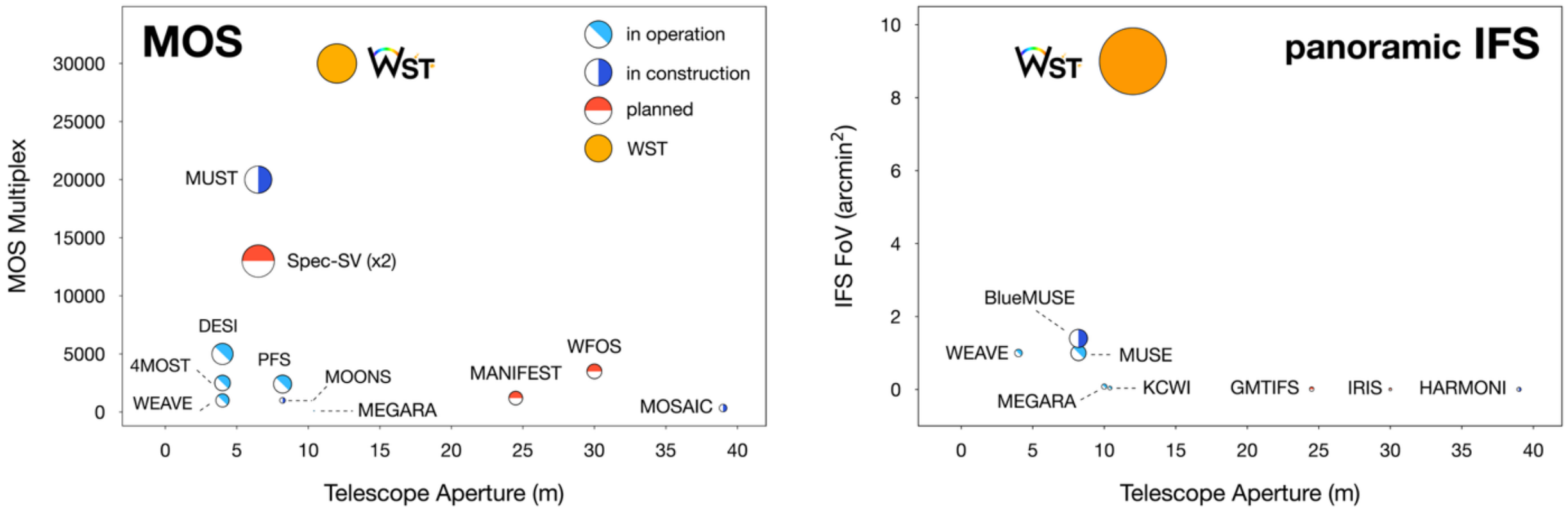

*Figure 1: Comparison of the WST MOS (left panel) and IFS (right panel) capabilities with existing and proposed ground-based spectroscopic facilities. Circle areas are proportional to the étendue (i.e., aperture times field of view area). For clarity, MOS or IFS with small multiplex or field of view are not shown in these figures. Multi-IFUs (e.g., KMOS, Hector or MaNGA) or sparse-field panoramic IFS (e.g., HETDEX) are not shown in this comparison, which only considers panoramic (monolithic) IFS with 100% fill-factor.*

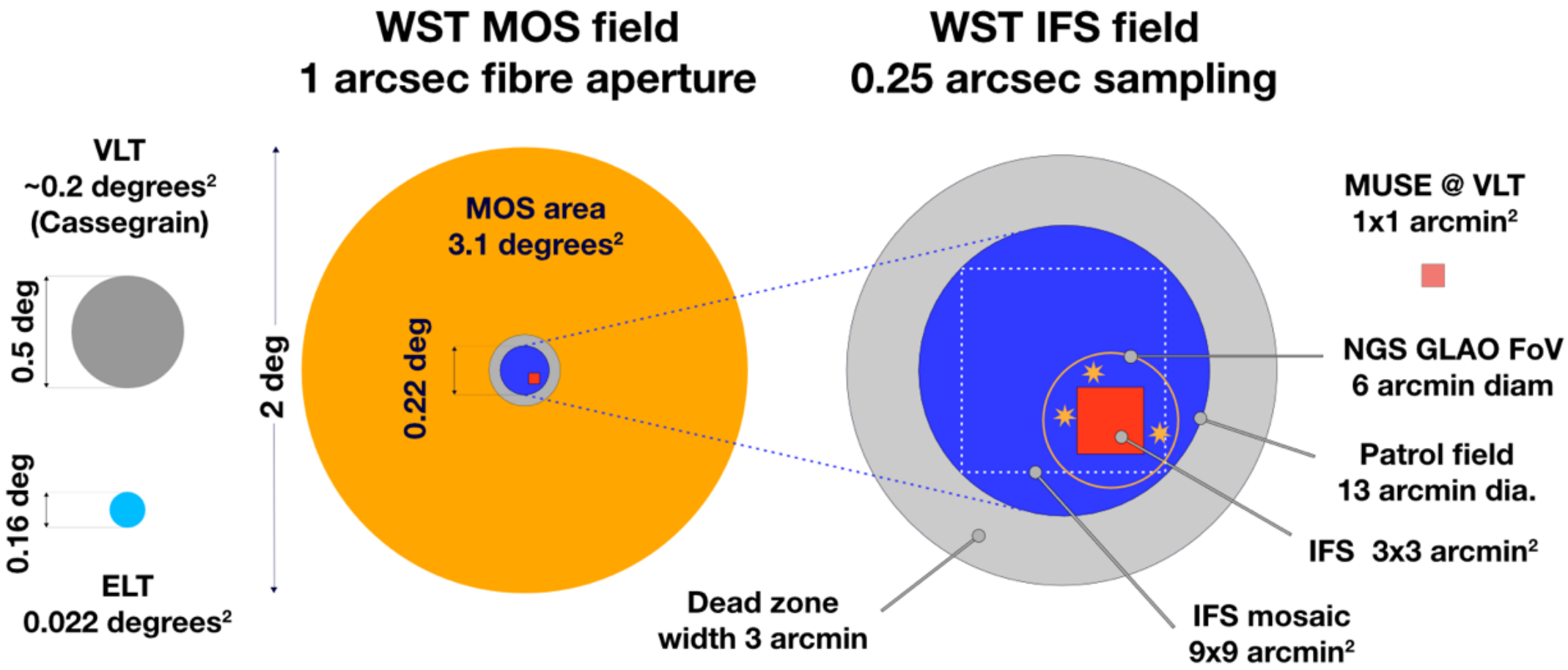

*Figure 2: The WST fields of view (FoV). The left panel shows the MOS FoV and the central circular area available for IFS observations. The right panel offers a closer view of the latter. The IFS 3x3 arcmin$^2$ FoV (in red) and its surrounding GLAO natural guide stars area (6 arcmin diameter) can be moved within the available area, providing the 9x9 arcmin$^2$ mosaic capability. There is a gap separating the MOS and IFS FoVs. The VLT, ELT and MUSE FoV are represented for comparison.*

## 3. SCIENCE CASE UPDATE

A preliminary version of the science white paper was published in 2024 [5]. Since then, the international science team, currently comprising 866 members (at the time of the paper submission), has undertaken a substantial effort to refine the scientific cases and identify those with the greatest relevance and highest transformative potential for the 2040s.

In response to the ESO Expanding Horizons call for short white papers in June 2025, the WST science team enthusiastically submitted 65 white papers highlighting the broad and transformative scientific impact that only such a facility could deliver. Although too numerous to discuss individually here, these contributions span all areas of astrophysics, from the Solar System to cosmology. Notably, many of the papers emphasize the strong synergies between the WST and current or future major astronomical facilities, including the ELT, Euclid, SKAO, Rubin Observatory/LSST, Roman Space Telescope, Einstein Telescope, LISA, and New Athena. This further reinforces the central role that the WST is expected to play within the global astronomical landscape of the 2040s.

Advances in the facility design have enabled the development of an exposure time calculator [6], allowing a quantitative assessment of instrument performance and supporting the optimization of survey design (see the Operations section).

In addition to the ESO short white papers, the consortium is preparing a series of more detailed white papers, which are planned to be published as a dedicated collection in the Open Journal of Astrophysics in 2027.

## 4. SYSTEM DESIGN

### 4.1 System description

The following table summarizes the main characteristics of the system:

| Item | Description |
|---|---|
| Telescope | 12-m segmented f/1 Cassegrain configuration<br>with dual focal plane operating in parallel:<br>- Seeing limited 2-deg dia. MOS focal plane at f/3 with ADC correction, accessible 370-1600 nm wavelength range.<br>- 6 arcmin dia. IFS focal plane over a 13 arcmin dia. patrol field with optical derotation and GLAO correction, 370-930 nm wavelength range. |
| Structure & Enclosure | Enclosure of 40-m diameter and 38-m height.<br>825 tons total & 130 tons altitude moving mass. |
| Adaptive Optics | IFS with ground-layer adaptive correction using natural guide stars and 1-m deformable mirror, 85% sky coverage at galactic pole and 0.5 arcsec FWHM at 650 nm with 50% probability. |
| Location | North Chile, in the Paranal-Armazones area. |
| Integral Field Spectrograph | 3x3 arcmin$^2$ with 0.25x0.25 arcsec$^2$ sampling, 370-930 nm simultaneous wavelength range.<br>192 integral field units composed of an image slicer and a two channels f/2.0 spectrograph (370-595 nm and 575-930 nm) with an average spectral resolution R~4800 in each band and two CMOS 6k x 6k 15 µm detectors. |
| | The IFS is located in the Coudé room and occupies a cylindrical envelope with a diameter of 26.4 m and a height of 6.8 m (3700 $m^3$). |
| Positioner | 32,000-unit robotic positioner for the 30,000 low-resolution spectrograph fibres and the 2,000 high-resolution spectrograph fibres. It places the 1 arcsec diameter fibres, which have a pitch of 6.2mm and a sub-7mm pitch depending on the modular architecture, to a precision of 5 microns RMS. Hexagonal field of view of 2-deg corner-to-corner covering an area of 2.6 $deg^2$. |
| Fibre sub-system | The fibre sub-system transports the light from the 32,000 positioners to the MOS-LR and MOS-HR spectrographs entrance slits. It is composed of connectors to positioner modules, Cassegrain cable wrap, elevation chain and fibre loop assemblies. |
| Multi Object Spectrograph Low Resolution mode | 54 units composed of f/1.6 spectrographs with 4 channels, 370-930 nm simultaneous wavelength range and 4 CMOS 6k x 6k 15 µm detectors. The first 3 channels achieve an average R~3800, while the redder channel has an average R~4900.<br>The spectrographs are located on the azimuth platform and occupy a total volume of ~170 $m^3$. |
| Multi Object Spectrograph High Resolution mode | 8 or 16 units composed of a fibre split (1 to 7 or 1 to 19) and a (f/1.5 or f/1.7) spectrograph with 4 channels covering the wavelength windows 401-443, 452-500, 542-600, 608-672 nm at R~40,000. The 7-fibre split version uses four 12k x 12k 10µm CMOS (either monolithic or a mosaic of four 6k x 6k CMOS), while the 19-fibre split design is based on 10k x 10k x 9 µm CMOS.<br>The spectrographs are located on the azimuth platform and occupy a total volume of ~200-400 $m^3$. |

### 4.2 Telescope system

Since the preliminary design presented in [3], several alternative configurations have been investigated and a trade-off analysis has been carried out to identify the preferred solution, balancing improved image quality with enhanced feasibility [7][8]. A key design innovation is the simultaneous delivery of two focal planes: a MOS focal plane providing a large 2-degree diameter FoV, and a central IFS focal plane. The latter features a 6-arcmin diameter that encompasses both the 3×3 $arcmin^2$ science field and the surrounding circular region used to acquire natural guide stars for the GLAO system. The IFS science field can be located anywhere within a 13 arcmin diameter patrol field.

The telescope optical design is presented in Fig. 3. The first part of the optical path is common to both MOS and IFS. It consists of M1, the 12-meter segmented primary mirror; M2, the 2.6-meter secondary tip-tilt mirror; and L1–L3, three movable 1.6-meter lenses forming the field corrector and atmospheric dispersion correction system. The beam is then focused at f/3.3 onto the MOS focal surface, where the fibre positioner is located. M1 comprises 78 segments of 1.4 m, closely based on the ELT segment design, differing primarily in their optical prescription.

It is worth noting that the WST MOS field of view will achieve an exceptional étendue of approximately 350 $m^2 \cdot deg^2$, about 50 times larger than that of the VLT focal plane (25 arcmin diameter) and around 10% greater than that of Rubin/LSST. This would place the WST among the highest-étendue ground-based optical facilities ever designed. Despite this exceptionally large étendue, the optical design maintains excellent image quality, achieving a RMS image quality of 0.13 to 0.30 arcsec at $z<30°$ all over the MOS 2 degree field of view.

The IFS optical path consists of a set of mirrors and lenses that relay the focal plane to the Coudé focus, where the IFS station is located. The pair M3-M4 selects and extracts the 6-arcmin field of view from the 13-arcmin patrol field centered on the telescope optical axis. Among the optical elements used to relay the beam, M7 is the 1-meter deformable mirror providing the GLAO correction. It is conveniently conjugated to an altitude of 100 m above the ground. An optical derotator (M11-M13 K-mirror) is used to compensate for field rotation, delivering a gravity-stable, non-rotating, telecentric focal plane at the entrance of the IFS station. The IFS optical train achieves excellent image quality, with an RMS image quality of 0.02-0.11 arcsec, depending on IFS subfield location within the patrol field. This stringent image-quality requirement arises from the improved point spread function delivered by the GLAO system.

The requirement to relay the IFS field of view to the Coudé focus without affecting the MOS focal plane implies the use of a large number of mirrors and lenses, and consequently a substantial number of optical surfaces (23). The resulting impact on the system throughput could potentially be significant enough to offset the benefit of the large 12-meter aperture. To mitigate this effect, we are developing advanced coatings for both mirrors and lenses [10]. The coatings used in the common optical path are designed to operate over the full wavelength range of the telescope (370–1600 nm), whereas the IFS optical path is restricted to the visible range (370–930 nm), allowing the use of coatings with higher efficiency. Thanks to these developments, the modeled telescope throughput is expected to reach average values of 79.8% and 91.3% for the IFS and MOS optical paths, respectively [10]. Even after accounting for a 10% manufacturing margin, the modeled average throughput of the WST IFS optical path reaches 72%, comparing remarkably well with the ~73% throughput of the VLT Nasmyth focus based on standard aluminum coatings, despite the substantially larger number of optical surfaces and the inclusion of an optical derotator in the WST IFS path.

The need to operate in parallel the MOS and IFS field of view requires a control strategy minimizing complexity, and above all avoiding cross-talk to the maximum possible extent. The MOS focal plane will drive the telescope, while the IFS optical path, which has many more degrees of freedom, will keep its focal plane aligned with that of the MOS system.

The telescope will be housed in a conventional alt-azimuth enclosure [9] inspired by the successful ESO VLT design. Ventilation louvers and a permeable wind screen will be used to minimize dome seeing and protect the telescope from excessive wind loading. Despite the large instrument complement, including nearly 200 integral-field spectrographs, the compact optical design based on an f/0.95 primary mirror allows the enclosure dimensions to remain moderate, with an estimated diameter of 40 m and a height of 38 m, only about 30% larger than those of the VLT enclosure.

The telescope structure (Fig. 4) has been designed to accommodate the substantial instrument mass associated with the MOS and IFS systems [9][18]. Current estimates indicate a total moving mass of approximately 825 tons, including around 225 tons allocated to the MOS spectrograph modules. Several configurations for the placement of these modules are under investigation, balancing fibre length, accessibility, structural performance, and seismic safety. In all cases, the modules are located behind the telescope structure to minimize their impact on airflow around the telescope.

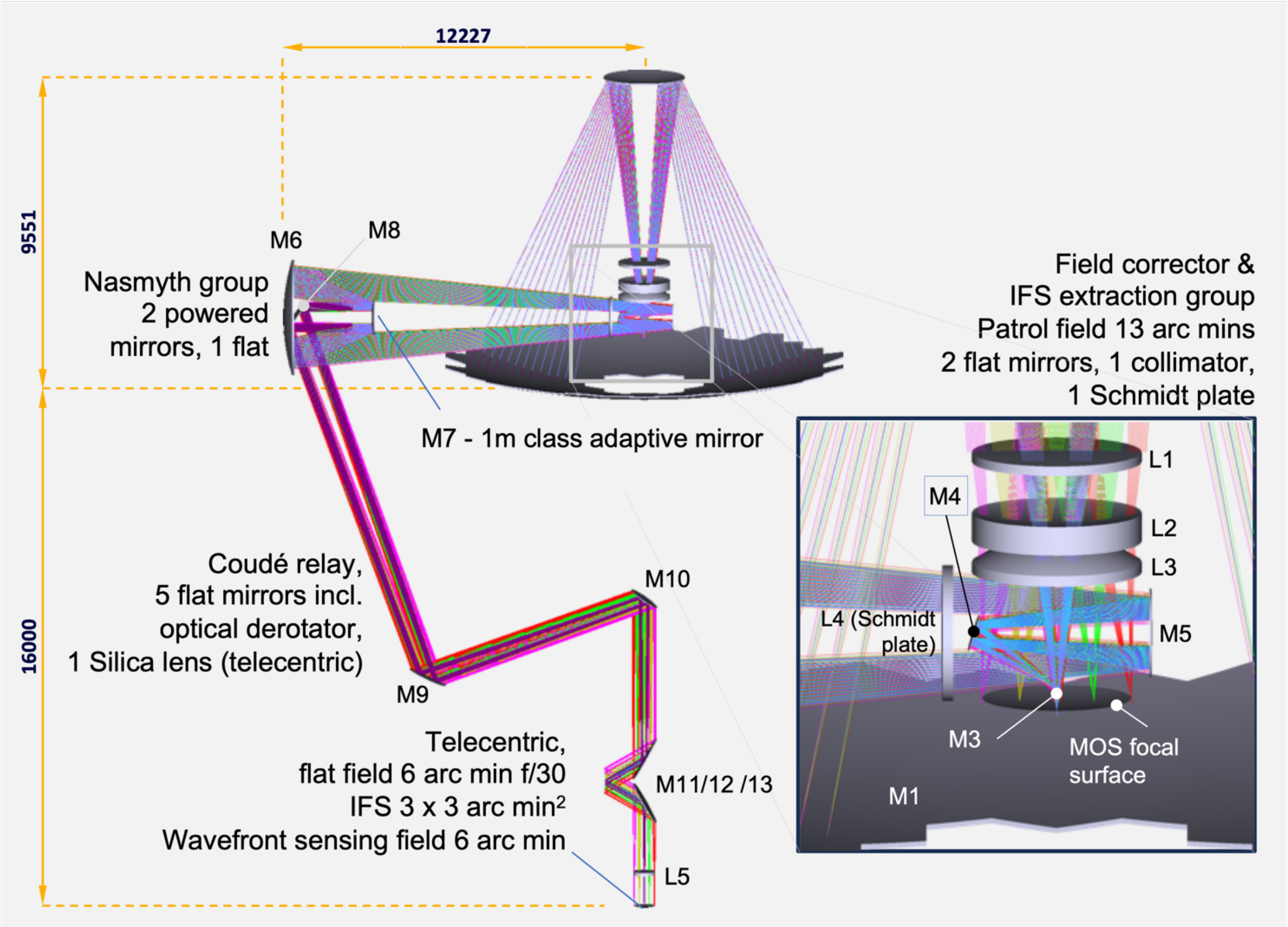

*Figure 3: Reference telescope optical design (all elements shown to scale).*

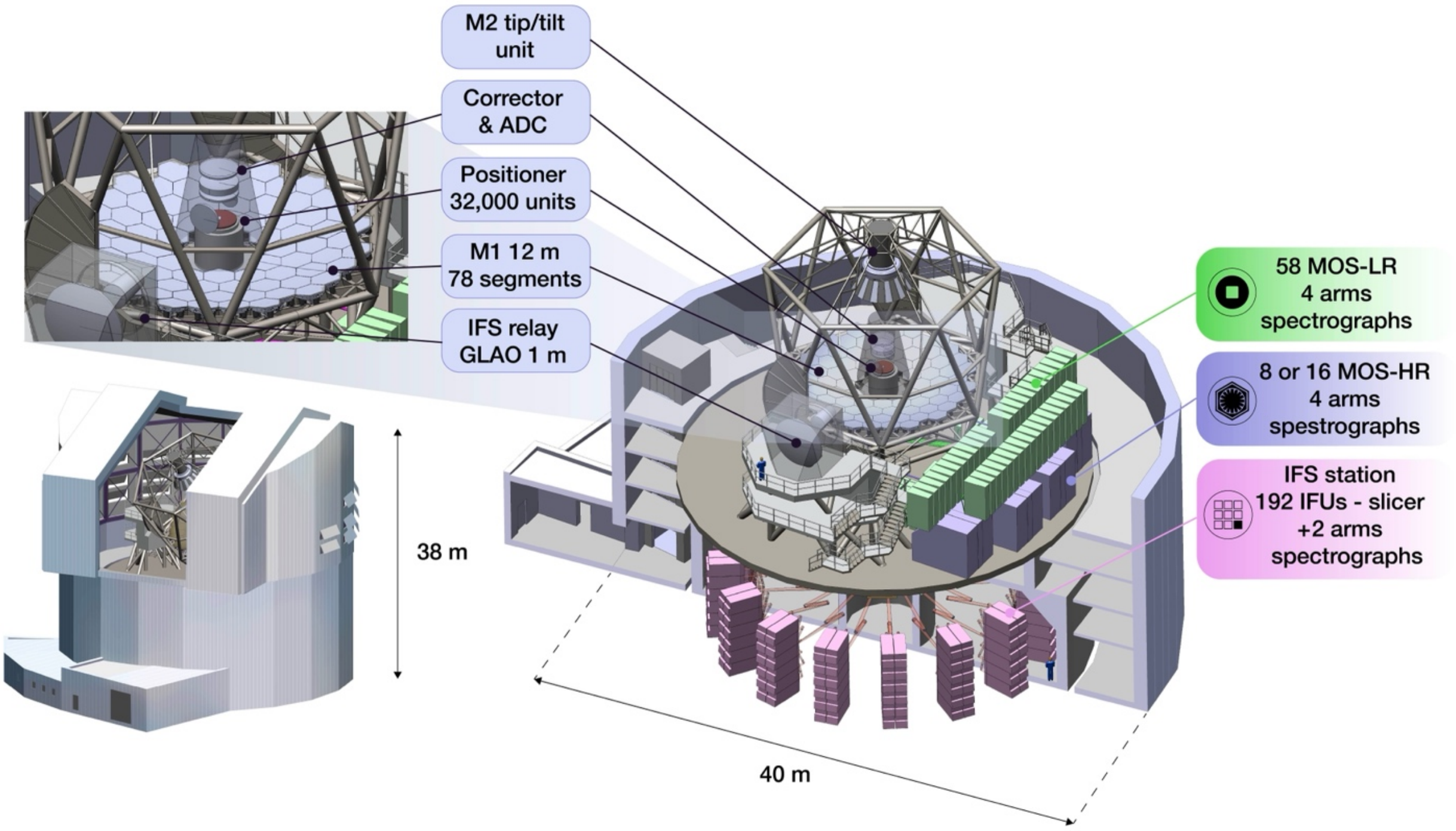

*Figure 4: Cutaway view showing the telescope and instruments inside the enclosure.*

### 4.3 Integral field spectrograph system

With the MOS system, the IFS represents a significant fraction of the overall facility volume and cost. The ambitious top-level requirements outlined in the first section — namely a 3 × 3 arcmin² field of view (nine times larger than MUSE), a spatial sampling of 0.25 arcsec (1.2 times coarser than MUSE), simultaneous wavelength coverage from 370 to 930 nm (combining the MUSE and BlueMUSE ranges), and an average spectral resolution of 4800 (1.7 times higher than MUSE) — immediately translate into very large detector real-estate requirements and, consequently, a large number of integral-field units, each comprising an image slicer, a spectrograph, and a detector system.

In MUSE, the 24 integral-field units already dominated both the cost and volume of the instrument [11]. With twice the étendue, nine times the field of view, and a broader spectral range, the design choices for the WST IFS became critically important. A straightforward extrapolation of the MUSE or BlueMUSE designs would have been risky, motivating a more comprehensive high-level design analysis.

In [12] we have systematically explored the WST IFS spectrograph design space across a broad range of detector formats, camera architectures, and spatial sampling options. The analysis consistently shows that, across nearly all metrics considered — including throughput, volume, cost, and construction carbon footprint — a larger number of simpler spectrographs is preferable to a smaller number of more complex units. This non-intuitive conclusion arises from the steep scaling of cost and volume with lens diameter, which outweighs the benefit of reducing the number of units.

The trade-off analysis led to the selection of a reference design comprising 192 integral-field units, each consisting of an image slicer, an f/2.0 two-channel spectrograph, 180 mm VPH gratings [13] and two 6k × 6k detectors with 15 µm pixels. We also explored the use of curved detectors [14] as a means to simplify the optical design, reduce cost, volume, and improve throughput. In particular, a variant of the reference design based on a cylindrically curved detector was considered in the trade study [12]. The detailed optical designs of both the reference and curved-detector variants are presented in [31]. The two spectrograph designs are shown in Fig. 5.

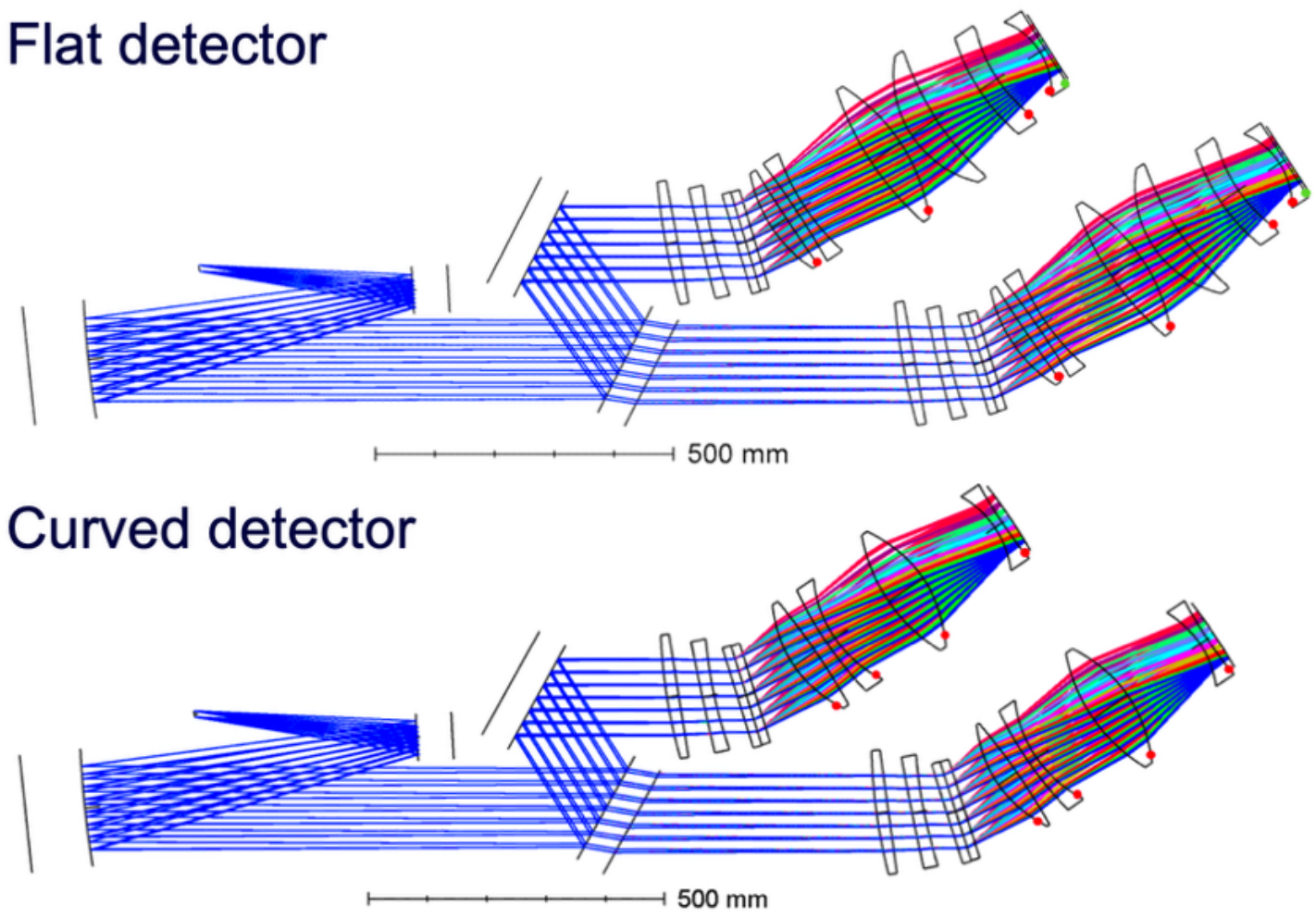

*Figure 5: The reference IFS spectrograph design (top) and its variant incorporating a curved detector (bottom). The benefits of the curved-detector design are clearly visible, including a reduced number of camera lenses (two fewer elements) and a more compact overall optical system.*

The image slicer is based on the Advanced Image Slicer concept [15], first developed for MUSE [11] and subsequently adopted as the baseline architecture for many modern integral field spectrographs, including KCWI, BlueMUSE, HARMONI, and MAVIS. However, the current manufacturing process is not well suited to the production of 192 image slicers within a reasonable cost and schedule. Alternative manufacturing approaches are therefore being investigated, including monolithic ultra-precision diamond machining [32] and laser-based shaping techniques [33], with the goal of enabling efficient large-scale production, while maintaining the required optical performance.

Calibration is performed using an internal calibration unit whose light can be injected directly into the IFS optical path. In addition, dome-flat calibrations are obtained using a dedicated deployable screen. The IFS station is enclosed and isolated from the observatory environment to minimize dust contamination and is equipped with active cooling to maintain a stable and uniform temperature. This controlled environment enhances the instrument's mechanical and thermal stability, thereby reducing the need for nighttime calibrations.

### 4.4 Multi-object spectrograph system

The multi-object spectrograph system is composed of three subsystems: the 32,000 fibre positioner, the low resolution spectrograph (MOS-LR) devoted to extreme multiplex (30,000), full optical spectral coverage (370-930 nm) at >3000 spectral resolution and the high resolution spectrograph (MOS-HR) with moderate multiplex (2,000) and a R~40,000 spectral resolution in a few spectral windows.

Each subsystem presents its own set of challenges. For the MOS-LR, the primary challenge, shared with the IFS, is the need to manufacture a large number of units to accommodate the required multiplex of 30,000 targets. We therefore adopted a similar design approach and explored a range of architectural options. Unsurprisingly, our study shows that a larger number of spectrographs equipped with simpler, slower cameras provides a more effective solution than a smaller number of spectrographs relying on very fast cameras with complex optical designs.

The selected architecture [16][17] comprises 54 low-resolution MOS spectrographs, each fed into four spectral channels and equipped with an f/1.64 camera (180-mm beam size). The detector format is common with the IFS, relying on 6k × 6k CMOS sensors with 15 µm pixels. The wavelength range from 370 to 930 nm is partitioned into four spectral bands: UB, V, R, and Iz. The design delivers a similar average spectral resolving power of R~3800 across the first three bands. The reddest band is designed to operate at a higher average resolving power, R~4900, providing improved discrimination and subtraction of the dense OH sky-line spectrum, thereby increasing observing efficiency at long wavelengths.

The key challenge for the MOS-HR subsystem is achieving a resolving power of R~40,000 on a telescope of this aperture. Our studies indicate that no practical solution exists without splitting the input light into multiple smaller fibres. This approach reduces the effective slit width seen by the spectrograph, allowing the size of the optical components to remain manageable. Two solutions are currently under investigation[23][24][25]: a design based on a 1-to-7 fibre split and an alternative based on a 1-to-19 fibre split. The final choice will depend on the trade-off between performance, complexity, and cost.

The baseline 1-to-7 fibre-split solution consists of eight spectrograph units, each incorporating four spectral channels, f/1.5 cameras, and a pupil diameter of 440 mm. The preferred detector is a 16k × 16k CMOS array with 10 µm pixels. As a fallback option, each focal plane can be implemented using a mosaic of four 6k × 6k CMOS detectors with the same pixel pitch. The four wavelength bands have been defined to maximize the scientific return by covering the spectral features required to measure the abundances of the key chemical elements targeted by the science case. The eight spectrograph units together occupy a total volume of approximately 441 $m^3$.

The alternative solution employs a 1-to-19 fibre split. Compared with the baseline architecture, this approach requires a larger number of spectrographs (16 instead of 8), but each spectrograph is significantly smaller and uses a smaller detector. Each unit comprises four spectral channels covering the same wavelength windows as the baseline design and is equipped with an f/1.7 camera with a pupil diameter of 180 mm and a 10k × 10k CMOS detector, featuring 9 µm pixels. The combined volume of the 16 spectrograph units is approximately 184 $m^3$.

While fibre splitting is essential to achieve the required resolving power, it introduces potential throughput losses that must be carefully controlled through the design of the splitting optics and fibre interfaces. In addition, around 2,000 splitter units will need to be produced and accurately aligned, posing significant manufacturing and integration challenges. A range of concepts is therefore being evaluated, with particular attention paid to throughput, alignment sensitivity, scalability, and cost.

The fibre-positioning system is one of the most challenging subsystems of the WST [19]. It must deploy 30,000 low-resolution fibres and 2,000 high-resolution fibres across the 3.1 $deg^2$ focal plane, while achieving a positioning accuracy of better than 5 μm RMS. This combination of multiplex, precision, and packing density exceeds the capabilities of current instruments by a substantial margin, and requires the development of a new generation of highly reliable robotic positioners.

The scale of the system presents an additional challenge, as more than 32,000 positioners will need to be manufactured, integrated, and operated efficiently. To mitigate the technical and industrial risks associated with such large-scale production, several positioner concepts are being pursued in parallel. Four architectures are currently under development [21][20]. Most concepts are based on modular triangular units with a 6.2 mm pitch and 63 positioners per module, while an innovative inline module design that produces identical modules to house 336 positioners is also being explored. Both module designs must accommodate an IFS mirror at the centre of the focal surface, introducing unique geometric and technical constraints that add further complexity to the design. The prototype systems are being evaluated in terms of positioning accuracy, repeatability, reconfiguration time, collision avoidance, reliability, and manufacturability.

The MOS calibration is performed with a dome screen (flat-fielding) and fibre light injection in the spectrograph (wavelength calibration).

## 5. OPERATIONS

The operations of the WST represent an important challenge for the project. Here, “operations” encompasses the entire survey lifecycle, from survey preparation and observation planning, to the telescope operation, data reduction, scientific analysis, and public data releases.

The operation of the telescope itself is expected to be relatively straightforward and has been designed to integrate seamlessly into the future operational framework of the Paranal Observatory. Apart from the MOS fibre-positioning system, the facility contains very few moving parts. In particular, the spectrographs do not require grating exchanges, simplifying both operations and maintenance. The GLAO system relies exclusively on natural guide stars and therefore avoids the complexity associated with laser guide star facilities. The few mechanisms in the optical path, such as the IFS field selector, MOS rotator and IFS derotator, are limited in number and, once commissioned, are expected to operate robustly with minimal maintenance requirements. In addition, advanced mirror coatings are being considered that would require only periodic cleaning rather than recoating. Such cleaning operations could be performed on site without removing the mirror segments, eliminating one of the most demanding maintenance activities faced by the ELT.

For the telescope operation, the most significant operational risk is associated with the fibre-positioning system and its 32,000 robotic actuators. A major effort is therefore underway to identify the optimal positioner architecture. While positioning accuracy, reconfiguration speed, and operational performance are important selection criteria, reliability will be equally critical. The mean time between failures (MTBF) of the full positioner system must be carefully assessed and will be a key parameter in the final trade-off, as the scientific productivity of the WST ultimately depends on the reliable operation of this unprecedented number of robotic devices.

As a dedicated survey telescope, the WST will operate differently from general-purpose observatories such as the VLT or the ELT. It is designed to conduct large spectroscopic surveys simultaneously with its three instruments: the IFS, the MOS-LR, and the MOS-HR (Fig. 6). Its operational model builds upon the successful approach adopted by 4MOST, where multiple surveys are carried out in parallel.

Compared to 4MOST, however, the WST introduces additional complexity through the inclusion of a large integral-field spectrograph operating simultaneously at the center of the field of view. The facility is also designed as a powerful time-domain instrument and must support efficient target-of-opportunity (ToO) observations, both through rapid fibre reconfiguration and telescope repointing.

These capabilities place stringent requirements on survey planning, scheduling, and operational flexibility. As part of the concept study, a prototype Facility Simulator code is being developed to address these challenges. The simulator ingests the requirements of all surveys and optimizes the scheduling of observations over a given time period according to the survey priorities to maximize the scientific return of each survey.

With 648 large format detectors, a single WST exposure generates approximately 54 GB of raw data per exposure, corresponding to about eight times the data volume of a Rubin/LSST exposure, roughly 60 times that of MUSE, and around 80 times that of 4MOST. Overall, the WST is expected to produce between 2 and 3 PB of raw data per year. In addition to handling this data volume, the facility must process observations rapidly enough to identify and characterize spectroscopic transients in near real time. The requirement to deliver homogeneous, science-ready advanced data products further increases the complexity of operations. Taken together, these characteristics firmly place the WST in the category of big-data astronomical facilities, alongside Rubin Observatory, SKAO, and major space-survey missions such as Euclid and Roman.

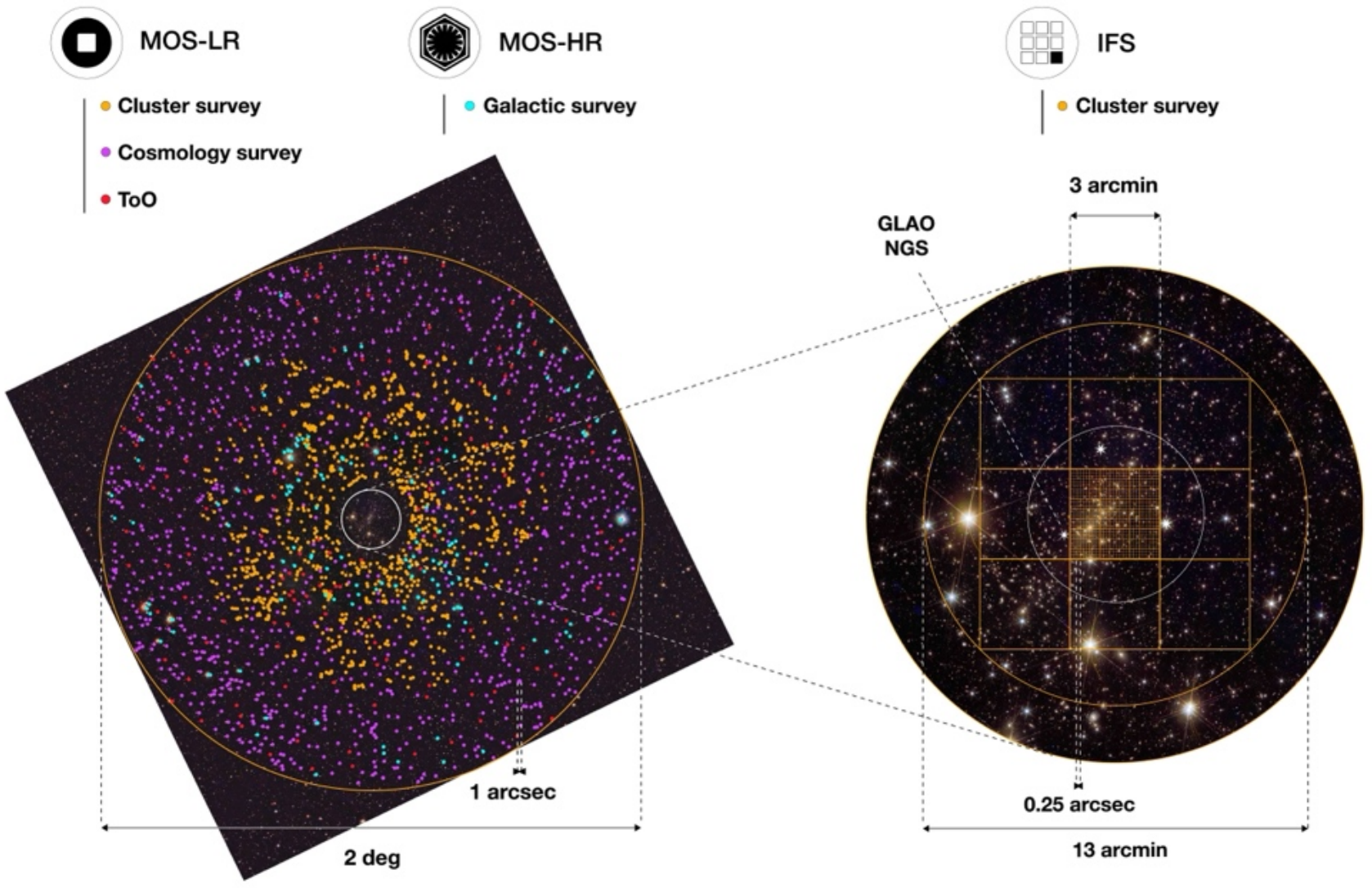

*Figure 6: The massive cluster survey is an example of multi-scale, multi-survey optimum use of the WST capabilities. The central region of the Abell 2390 cluster will be imaged by the IFS, whose field of view is ideally suited to probing the strong lensing region, either with a single 3×3 arcmin² pointing (700 kpc at the cluster distance) or through a mosaic of IFS observations. Sufficiently bright stars are present within the surrounding 6 arcmin diameter technical field of view to enable effective GLAO correction, delivering superb image quality across the IFS field. At the same time, a fraction of the 30,000 MOS-LR fibres can be dedicated to tracing the distribution of cluster galaxies on Mpc scales, providing key insights into the cosmic web structure surrounding the massive cluster. The remaining MOS-LR fibres can be used in parallel for large-scale cosmology surveys, while MOS-HR fibres will measure chemical abundances in Milky Way halo stars. Finally, alerts from photometric time-domain surveys (e.g., LSST) will enable the observation of targets of opportunity (ToO), using a pre-allocated subset of MOS-LR fibres. Credits: M-C Fortuna and R. Bacon. The Abell 2390 central image is from Euclid: ESA/Euclid/Euclid Consortium/NASA, image processing by J.-C. Cuillandre (CEA Paris-Saclay) and G. Anselmi. The large scale 2-degree image centered on the cluster is from Pan-Starrs (Pan-STARRS1 Surveys (PS1), Institute for Astronomy, University of Hawaii).*

## 6. SUSTAINABILITY

The environmental impact has been considered as an integral part of the WST concept study from the earliest design phases. We have performed an initial life-cycle assessment of the main instrument architectures, detector technologies, cooling systems, and operational scenarios in order to quantify their environmental impact and identify potential mitigation strategies [27]. The results show that environmental considerations can be incorporated into the design trade-offs without compromising scientific performance. In particular, architectures based on CMOS detectors, FPGA-based electronics, and shared $CO_2$ cooling systems offer substantial reductions in both energy consumption and carbon footprint compared to more traditional CCD-based solutions. Note also that, as part of the Paranal-Armazones operation, the WST will benefit from all the other sustainability changes ESO is making there.

The scale of WST operations, discussed in the previous section, also raises important sustainability questions related to data management. As a big-data facility, the WST will require significant computing, storage, and networking resources throughout its operational lifetime. We have therefore begun to assess the environmental impact associated with data transfer, processing, and long-term storage, and to investigate strategies that minimise these impacts while preserving the rapid processing capabilities and high-quality data products required by the science programme. These studies demonstrate that sustainability considerations should extend beyond the hardware design and be incorporated into the overall operational concept of the facility.

Beyond hardware design and data management, the scale of WST operations also implies important societal sustainability challenges related to governance, collaboration, and team management across a large, international, and interdisciplinary community. To address these aspects, we have established an Equity, Diversity, and Inclusion working group within the Horizon Europe conceptual study. This group supports inclusive governance and communication practices, broadens participation across the Collaboration, and develops shared community infrastructure, including the WST Code of Conduct and associated accountability mechanisms. These activities recognise that the long-term sustainability of a major scientific facility also depends on the resilience, accessibility, and inclusiveness of its community. The activities and first lessons from the WST EDI working group are discussed in more detail in a companion SPIE contribution [30].

## 7. KEY TECHNICAL CHALLENGES AND DEVELOPMENT ROADMAP

For many subsystems, the facility design relies on proven, technologies with a high technology readiness level (TRL), thereby minimizing technical risk. A notable example is the 12-m segmented primary mirror, which is based directly on the ELT segment design, differing only in its optical prescription. The altitude structure incorporates a segment-handling bridge similar to that adopted for the ELT, and the project foresees reusing much of the existing ELT segment maintenance infrastructure. Given that the WST primary mirror comprises only about 10% of the ELT segment population, this approach is expected to provide an efficient and cost-effective solution for segment integration, maintenance, and phasing operations. By leveraging ESO's substantial investment in the development and qualification of these state-of-the-art opto-mechanical units, the WST can incorporate mature and well-characterized components into its design, significantly reducing development costs, schedule risks, and technical uncertainties.

For other subsystems, no existing technology was found to fully meet the ambitious WST top-level requirements, and innovative solutions have therefore been developed. Examples include the fibre-positioning system, curved detectors, and advanced optical coatings. These technologies offer the potential for substantial gains in performance, cost, or operational efficiency, but require dedicated development to reach the maturity needed for deployment in a facility of this scale. At least some of these technologies will have wider use in other ESO instruments and facilities, offering two-way synergies.

The maturity levels of these developments vary. Some concepts are still at relatively low TRL and have entered an active prototyping phase, such as the FLEX fibre-positioner architecture [21] and graded-index anti-reflection coatings [10]. Others, such as curved detectors, have benefited from several years of development and demonstrated promising results [28], but have not yet reached the level of industrial maturity required for large-scale production. For these technologies, dedicated prototype programmes have been initiated, in collaboration with industry, to assess their manufacturability, performance, reliability, and readiness for incorporation into the final facility design.

These developments inevitably carry a degree of technical risk. To manage this risk, a dedicated technology roadmap is being developed to identify the activities, milestones, and resources required to mature each technology to the TRL needed for the construction phase. This roadmap includes prototype development, performance validation, industrialization studies, and reliability assessments, allowing risks to be identified and mitigated at an early stage.

In parallel, backup solutions based on mature, high-TRL technologies have been identified for all critical developments. This dual-path strategy ensures that the facility requirements can still be met should one of the innovative solutions fail to reach the required level of performance or industrial readiness. A good example is the use of curved detectors, which offer significant benefits in terms of optical simplification and cost reduction. Should the industrialization of the detector-curving process prove unable to meet the required performance, yield, or production rate, the corresponding spectrographs can revert to a design based on conventional flat detectors, albeit with some loss in overall system optimization.

Another important challenge is the large-scale production required for several key subsystems, including approximately 192 image slicers, 260 spectrographs, and 648 detectors. These quantities are substantial by the standards of optical astronomical instrumentation, with only a handful of projects having involved production at a similar scale, such as VIRUS with 156 spectrographs [34] or LSST with 189 science-grade CCDs [35]. Consequently, the concept study has devoted significant effort to identifying designs and technologies that can be produced efficiently at scale without compromising performance.

For the detectors, CMOS technology has been selected as the baseline solution, with CCDs retained as a backup option. This choice is motivated by several factors, including lower environmental impact, reduced operational costs, and the intrinsic advantages of CMOS architectures, such as fast and non-destructive readout capabilities. In addition, the commercial semiconductor industry has demonstrated the ability to manufacture large quantities of CMOS sensors at affordable cost. While many of the key performance requirements of WST can already be met by recent CMOS developments [29], no commercial CMOS sensor currently satisfies the full set of specifications required by the facility. However, given the large number of detectors needed for WST, the development of a custom CMOS sensor is considered both feasible and economically justified. The associated non-recurring engineering costs remain modest when compared to the overall cost of the detector procurement and the facility as a whole.

The manufacture of approximately 260 high-performance spectrographs represents one of the most significant challenges of the WST project, as it is expected to be a major cost driver for the facility. A spectrograph is a complex opto-mechanical system requiring the fabrication, integration, and alignment of numerous components to tight tolerances. While the community has gained valuable experience through the successful construction of the 24 MUSE spectrographs, the WST requires more than an order of magnitude increase in production scale. Manufacturing and alignment approaches that are practical for a handful of units, or even a few dozen, may become prohibitively expensive, time-consuming, or difficult to industrialize when scaled to hundreds of units.

To address this challenge, we have initiated discussions with industrial partners at an early stage of the concept study to investigate manufacturing and assembly processes capable of delivering such a large number of spectrographs within realistic cost and schedule constraints. These interactions provide valuable feedback that can be incorporated directly into the design process, allowing the instrument architecture to be optimized for serial production, assembly, testing, and quality control. This approach has already influenced several key design decisions. For example, trade-off studies for the IFS spectrographs have shown that a larger number of simpler spectrographs can be more economical and easier to industrialize than a smaller number of highly complex units [12]. Such considerations are becoming an integral part of the overall design strategy and are helping to ensure that the facility remains both technically feasible and economically viable.

The big-data nature of the WST represents a significant challenge that must be anticipated and addressed well in advance of operations. The facility will generate unprecedented volumes of spectroscopic data for an ESO observatory and will require efficient solutions for data transport, storage, processing, quality control, transient identification, and public data releases. Developing the necessary infrastructure, software, and operational procedures will therefore be a major component of the project.

Fortunately, although ESO does not yet operate a facility with data rates comparable to those expected for the WST, the European astronomical community is already deeply involved in several large-scale data-intensive projects, including Gaia, Rubin Observatory/LSST, SKAO, Euclid, and Roman. The WST can therefore benefit from a substantial body of expertise in survey operations, distributed computing, data management, and advanced data-product generation. This creates an excellent opportunity for knowledge transfer and collaboration. We therefore envisage that the operational data challenges of the WST, including large-scale processing exercises and end-to-end data challenges, will be developed jointly by ESO and the relevant scientific and technical communities, allowing experience from these major facilities to be leveraged from the earliest phases of the project.

It is important to note that the challenge is not limited to the sheer volume of data that must be processed. The complexity of the data products is equally important. The WST will not simply produce images, but vast quantities of spectra and integral-field datacubes. The latter are particularly demanding because they combine imaging and spectroscopy over large fields, enabling the detection of unexpected phenomena and serendipitous discoveries that may not be captured by predefined analysis pipelines. Fully exploiting this discovery potential will require the development of advanced processing, analysis, and data-mining techniques capable of extracting scientific value beyond the original survey objectives.

Fortunately, the ESO community has already accumulated significant expertise in handling complex spectroscopic datasets through facilities such as MUSE, and soon 4MOST, WEAVE, MOONS and BlueMUSE. While these instruments operate at a much smaller scale than WST, they provide invaluable testbeds for developing and validating the methodologies, software frameworks, and science pipelines that will ultimately be required. Building on this experience is essential to reduce operational risk and to ensure that the data-processing ecosystem, advanced data products, and scientific exploitation tools are mature and fully operational when the WST begins observations.

An important aspect considered during the concept study is the long-term evolution of the facility. For general-purpose observatories, scientific evolution is typically achieved through successive generations of instruments. For a dedicated survey facility such as WST, the situation is less straightforward. The telescope and instruments are being designed as an integrated system, optimized to meet a well-defined set of scientific objectives. While this approach maximizes performance and efficiency, it can also reduce flexibility, as some design choices may constrain future upgrades or limit the introduction of new capabilities.

To address this issue, we are investigating several possible evolutionary paths for WST beyond its initial science programme. Among the options under consideration are an extension of the MOS capabilities into the near-infrared, potentially reaching wavelengths of up to 1600 nm, and the replacement of single-object fibres in the MOS focal plane by mini-IFUs. Such developments would significantly expand the scientific reach of the facility, enabling new studies of galaxy evolution, stellar populations, and the high-redshift Universe. By identifying these opportunities at an early stage, we aim to ensure that the baseline design preserves sufficient flexibility to accommodate future upgrades and to maximize the scientific impact and longevity of WST over several decades.

## 8. CONCLUSIONS

The WST concept study has demonstrated that a dedicated wide-field spectroscopic survey facility combining extreme-multiplex multi-object spectroscopy and panoramic integral-field spectroscopy on a 12-m class telescope is both scientifically compelling and technically credible. Supported by Horizon Europe, the study has refined the science drivers and top-level requirements, explored a wide range of architectural options, and established reference designs for the telescope, instruments, operations, sustainability framework, and technology roadmap. Together, these activities provide a solid foundation for the next phases of development.

The WST occupies a unique and currently unmatched position among future astronomical facilities. No existing or planned observatory combines a 30,000-object low-resolution spectroscopic survey capability, a 2,000-object high-resolution spectroscopic survey capability, and a panoramic integral-field spectrograph operating simultaneously on a 12-m telescope. This combination will enable transformational science across all areas of astrophysics, from the Solar System and Galactic archaeology to galaxy evolution, cosmology, time-domain astronomy, and multi-messenger astrophysics.

The study has also identified the principal challenges associated with the facility, including large-scale instrument replication, advanced detector technologies, massive data processing, and the operation of an unprecedented fibre-positioning system. Importantly, credible mitigation strategies and technology-development paths have been defined for all major risk areas. The WST builds on a strong European heritage in wide-field spectroscopy and integral-field astronomy, benefiting from the expertise gained through projects such as MUSE, 4MOST, MOONS, Gaia, Euclid, Rubin, and SKAO, while extending these capabilities by more than an order of magnitude in several key areas.

Beyond its baseline design, the WST has been conceived with the ambition to remain scientifically competitive for decades. Future developments such as near-infrared spectroscopic capabilities or mini-IFU deployment offers promising paths for extending its scientific reach and maximizing its long-term impact.

As astronomy enters an era dominated by large imaging, time-domain, gravitational-wave, and multi-messenger surveys, the need for equally powerful spectroscopic facilities is becoming increasingly evident. The WST is designed to provide this missing capability. By delivering the spectroscopic information required to interpret, calibrate, and fully exploit the discoveries of the major facilities of the 2040s, the WST will be a flagship astronomical infrastructure in the post-ELT construction era and a cornerstone of observational astrophysics for decades to come.

*We dedicate this paper to the memory of Bianca Garilli, who passed away in 2024. Bianca was an active member of the Project Office and was deeply involved in the WST project. Her enthusiasm, dedication, and contributions played an important role in advancing the project, and she is deeply missed by her colleagues and collaborators.*

Acknowledgments: This project has received funding from the European Union's Horizon Europe research and innovation programme under grant agreement No. 101183153. INAF authors acknowledge funding from the INAF 2023 Large Grant WST: the Wide-field spectroscopic telescope